\documentclass[a4paper]{article}
\usepackage{graphicx}
\usepackage{amsmath}

\begin{document}

\title{The entanglement of damped noon-state and its performance in phase
measurement \\
}
\date{}
\author{Xiao-Yu Chen, Li-zhen Jiang, Liang Han \\
Lab. of quantum Information, China Institute of Metrology}
\maketitle

\begin{abstract}
The state evolution of the initial optical \textit{noon} state is
investigated. The residue entanglement of the state is calculated after it
is damped by amplitude and phase damping. The relative entropy of
entanglement of the damped state is exactly obtained. The performance of
direct application of the damped \textit{noon} state is compared with that
of firstly distilling the docoherence damped state then applying it in
measurement.
\end{abstract}

\section{Introduction}

Quantum entanglement between two or more particles has attracted great
interest and produced many applications in quantum information processing,
such as quantum communication,quantum computation and quantum cryptography
and quantum metrology. It has been known for some time that entangled states
can be used to perform supersensitive measurements, for example in optical
interferometry or atomic spectroscopy\cite{Holland} \cite{Bollinger} \cite
{Dowling}. The idea has been demonstrated for entangled states of two photons%
\cite{DAngelo}, three photons \cite{Mitchell} and four photons \cite{Walther}%
. In the best case, the interferometric sensitivity can reach the quantum
mechanical 'Heisenberg-limit' in entanglement enhanced measurement which
overwhelms the classical shot noise limit. If $\phi $ is the phase to be
estimated, and $N$ is the number of independent trials in the estimation,
the classical shot noise limit is $\Delta \phi =1/\sqrt{N}.$ When entangled
state is used, the limit can be reduced to at most to 'Heisenberg-limit' $%
\Delta \phi =1/N$. One of such entangled states is the so-called \textit{noon%
} state
\begin{equation}
\left| N::0\right\rangle _{ab}=\frac 1{\sqrt{2}}(\left| N,0\right\rangle
_{ab}+\left| 0,N\right\rangle _{ab})
\end{equation}
which describes two modes $a,b$ in a superposition of distinct Fock states $%
\left| n_a=N,n_b=0\right\rangle $ and $\left| n_a=0,n_b=N\right\rangle .$
The applications of this state include quantum metrology \cite{Holland} \cite
{Bollinger} \cite{Dowling} \cite{Campos} and quantum lithography \cite{Boto}%
. In all the applications of \textit{noon} state, the decoherence of the
state and the performance of the damped state are less concerned. Huelga
\textit{et al} considered the ion system in presence of decoherence\cite
{Huelga}. It is inevitable that quantum state interacts with environment
which will cause the decoherence of the state, thus in quantum
supersensitive measurements decoherence should be included. We in this paper
will investigate the entanglement of optical \textit{noon }state in presence
of decoherence and the performance degradation.

\section{Decoherence}

A quantum state will undergo decoherence after preparation. The decoherece
comes from the interaction with environment. For continuous variable (CV)
system, two most popular decoherences are amplitude damping and phase
damping. The master equation describing these two decoherences for the
density operator $\rho $ is \cite{Kinsler} \cite{Lindblad}\cite{Walls} (in
the interaction picture) $\frac{d\rho }{dt}=(\mathcal{L}_1\mathcal{+L}%
_2)\rho ,$with $\mathcal{L}_1$ represents the amplitude damping concerning
with vacuum environment,
\begin{equation}
\mathcal{L}_1\rho =\sum_i\frac{\Gamma _i}2(2a_i\rho a_i^{\dagger
}-a_i^{\dagger }a_i\rho -\rho a_i^{\dagger }a_i),
\end{equation}
and $\mathcal{L}_2$ represents the phase damping,
\begin{equation}
\mathcal{L}_2\rho =\sum_i\frac{\gamma _i}2[2a_i^{\dagger }a_i\rho
a_i^{\dagger }a_i-(a_i^{\dagger }a_i)^2\rho -\rho (a_i^{\dagger }a_i)^2],
\end{equation}
with $a_i$ the annihilation operation of $i-th$ mode, and $\Gamma _i$ and $%
\gamma _i$ are damping coefficients of $i-th$ mode for amplitude and phase
damping respectively. The solution to the master equation can be
conveniently obtained by first transforming the equation to the diffusion
equation of the characteristic function of the state, then solve the the
differential equation of the characteristic function. The time evolution
solution of he density operator can be recovered from the characteristic
function.

The characteristic function is defined as $\chi =tr[\rho \mathcal{D}(\mu )]$%
, where $\mathcal{D}(\mu )=\exp (\mu a^{\dagger }-\mu ^{*}a)$ is the
displacement operator, with $\mu =[\mu _1,\mu _2,\cdots ,\mu _s]$ $%
,a=[a_1,a_2,\cdots ,a_s]^T$ and the total number of modes is $s$. The
diffusion eqation of the characteristic function will be \cite{Chen}
\begin{equation}
\frac{\partial \chi }{\partial t}=-\frac 12\sum_j\Gamma _j\{\left| \mu
_j\right| \frac{\partial \chi }{\partial \left| \mu _j\right| }+\left| \mu
_j\right| ^2)\chi \}+\frac 12\sum_j\gamma _j\frac{\partial ^2\chi }{\partial
\theta _j^2},
\end{equation}
where $\mu _j=\left| \mu _j\right| e^{i\theta _j}.$ The solution is simply
be
\begin{eqnarray}
\chi (\mu ,\mu ^{*},t) &=&\int dx\chi (\mu e^{-\frac{\Gamma t}2+ix},\mu
^{*}e^{-\frac{\Gamma t}2-ix},0)\prod_j(2\pi \gamma _jt)^{-1/2} \\
&&\exp [-\frac{x_j^2}{2\gamma _jt}-\frac 12(1-e^{-\Gamma _it})\left| \mu
_j\right| ^2].  \nonumber
\end{eqnarray}
with $\mu e^{-\frac{\Gamma t}2+ix}$ standing for $(\mu _1e^{-\frac{\Gamma _1t%
}2+ix_1},\mu _2e^{-\frac{\Gamma _2t}2+ix_2},\cdots ,\mu _se^{-\frac{\Gamma
_st}2+ix_s})$. The time dependent state can be recovered by $\rho =\int
[\prod_j\frac{d^2\mu _j}\pi ]\chi (\mu ,\mu ^{*},t)D(-\mu )$ \cite{Perelomov}%
.

The characteristic function of the \textit{noon} state is
\begin{eqnarray}
\chi (\left| N::0\right\rangle ) &=&\left\langle N::0\right| D(\mu )\left|
N::0\right\rangle  \\
&=&\frac 12e^{-\frac{\left| \mu \right| ^2}2}[L_N(\left| \mu _1\right|
^2)+L_N(\left| \mu _2\right| ^2)+\frac 1{N!}((-\mu _1^{*}\mu _2)^N+(-\mu
_1\mu _2^{*})^N)].  \nonumber
\end{eqnarray}
where $L_N$ is the Laguerre polynomial of order $N:$ $L_N(z)=\sum_{m=0}^N%
\frac{(-z)^m}{m!}\binom Nm.$ The time evolution of the characteristic
function will be
\begin{eqnarray}
\chi (\mu ,\mu ^{*},t) &=&\frac 12e^{-\frac{\left| \mu \right| ^2}%
2}[L_N(\left| \mu _1\right| ^2e^{-\Gamma _1t})+L_N(\left| \mu _2\right|
^2e^{-\Gamma _2t}) \\
&&+\frac 1{N!}e^{-N\overline{\Gamma }t-N^2\overline{\gamma }t}((-\mu
_1^{*}\mu _2)^N+(-\mu _1\mu _2^{*})^N)],  \nonumber
\end{eqnarray}
where $\overline{\Gamma }=\frac 12(\Gamma _1+\Gamma _2),$ $\overline{\gamma }%
=\frac 12(\gamma _1+\gamma _2).$ The solution to the master equation of
density operator will be
\begin{eqnarray}
\rho  &=&\frac 12\{\sum_{m=0}^N\binom Nm[(1-e^{-\Gamma
_1t})^{N-m}e^{-m\Gamma _1t}\left| m0\right\rangle \left\langle m0\right|  \\
&&+(1-e^{-\Gamma _2t})^{N-m}e^{-m\Gamma _2t}\left| 0m\right\rangle
\left\langle 0m\right| ]  \nonumber \\
&&+e^{-N\overline{\Gamma }t-N^2\overline{\gamma }t}[\left| N0\right\rangle
\left\langle 0N\right| +\left| 0N\right\rangle \left\langle N0\right| ]\}.
\nonumber
\end{eqnarray}
The integral on $\mu $ is carried out by the technique of integral within
ordered operators.

\section{The entanglement of the damped state}

The damped state $\rho $ is a mixed state. According to Peres-Horodecki
criterion, the state is always entangled. The entanglement of the state can
be carried out if measured by relative entropy of entanglement. The relative
entropy of $\rho $ with respect to a separable state $\sigma $ is $S(\rho
\left\| \sigma \right. )=Tr(\rho \log _2\rho -\rho \log _2\sigma )$, the
relative entropy of entanglement of $\rho $ is the minimization of $S(\rho
\left\| \sigma \right. )$ over all separable state $\sigma .$ Let the
extremal separable state that minimizes the relative entropy be $\sigma ^{*}$%
. Denote $\rho =c_{00}\left| 00\right\rangle \left\langle 00\right|
+\sum_{m=1}^N(c_{m0}\left| m0\right\rangle \left\langle m0\right| +$ $%
c_{0m}\left| 0m\right\rangle \left\langle 0m\right| )$ $+c(\left|
N0\right\rangle \left\langle 0N\right| +\left| 0N\right\rangle \left\langle
N0\right| )$. We will obtain $\sigma ^{*}$ by first suppose $\sigma ^{*}$
having a special form then prove that $\sigma ^{*}$ is extremal (see
Appendix) . Suppose
\begin{eqnarray}
\sigma ^{*} &=&d_{00}\left| 00\right\rangle \left\langle 00\right|
+\sum_{m=1}^N(d_{m0}\left| m0\right\rangle \left\langle m0\right|
+d_{0m}\left| 0m\right\rangle \left\langle 0m\right| ) \\
&&+d(\left| N0\right\rangle \left\langle 0N\right| +\left| 0N\right\rangle
\left\langle N0\right| )+d_{NN}\left| NN\right\rangle \left\langle NN\right|
.  \nonumber
\end{eqnarray}
The function that should be minimized is
\begin{eqnarray}
-Tr(\rho \log _2\sigma ^{*}) &=&-c_{00}\log
_2d_{00}-\sum_{m=1}^{N-1}(c_{m0}\log _2d_{m0}+c_{m0}\log _2d_{m0}) \\
&&-Tr\left[
\begin{array}{ll}
c_{N0} & c \\
c & c_{0N}
\end{array}
\right] \log _2\left[
\begin{array}{ll}
d_{N0} & d \\
d & d_{0N}
\end{array}
\right] .  \nonumber
\end{eqnarray}
The constraints are $Tr\sigma ^{*}=1$ and $d^2=d_{00}d_{NN}$ , the later
specifies that the extremal separable state should be at the edge of the
separable state set (e.g.\cite{Chen0}). For the general situation of $%
c_{N0}\neq c_{0N},$ the minimization problem have not an analytical solution%
\cite{Chen1}. When the amplitude damping is symmetric, we have $\Gamma
_1=\Gamma _2=\Gamma $ $($thus $c_{m0}=c_{0m};$ $m=1,\cdots ,N),$ the
solution to the minimization problem is
\begin{eqnarray}
d_{m0} &=&d_{0m}=c_{0m,}\text{ for }m=1,\cdots ,N-1; \\
d_{00} &=&\frac{(c_{N0}+c_{00})^2c_{00}}{(c_{N0}+c_{00})^2-c^2};\text{ }%
d_{NN}=\frac{c^2c_{00}}{(c_{N0}+c_{00})^2-c^2}; \\
d_{0N} &=&d_{N0}=c_{N0}-d_{NN};\text{ }d=\frac{c(c_{N0}+c_{00})c_{00}}{%
(c_{N0}+c_{00})^2-c^2};
\end{eqnarray}
The relative entropy of entanglement of state $\rho $ is:\ $E_r(\rho )$ $%
=Tr\rho (\log _2\rho -\log _2\sigma ^{*})=c_{00}\log \frac{c_{00}}{d_{00}}%
+c_{+}\log \frac{c_{+}}{d_{+}}$ $+c_{-}\log \frac{c_{-}}{d_{-}},$ with $%
c_{\pm }=c_{N0}\pm c,$ $d_{\pm }=d_{N0}\pm d$. It can be written as
\begin{equation}
E_r(\rho )=2(c_{00}+c_{N0})[1-H_2(\frac{c_{00}+c_{N0}+c}{2(c_{00}+c_{N0})})],
\end{equation}
where $H_2(\varepsilon )=-\varepsilon \log _2\varepsilon -(1-\varepsilon
)\log _2(1-\varepsilon )$ is the binary entropy function, and $%
c_{00}=(1-e^{-\Gamma t})^N,$ $c_{N0}=\frac 12e^{-N\Gamma t},c=\frac
12e^{-N\Gamma t-N^2\overline{\gamma }t}$. It should be mentioned that the
corresponding solution to the problem of two qubits system is known \cite
{Vedral} \cite{Chen1}.

Other entanglement measures are entanglement of formation and distillable
entanglement. From the definition of the entanglement of formation, it is
easily to obtained an upper bound for the entanglement of formation, which
is
\begin{equation}
E_f^{+}(\rho )=(c_{N0}+c_{0N})H_2(\frac{1+\sqrt{1-c^2/(c_{N0}+c_{0N})^2}}2).
\end{equation}
We suspect if this is just the entanglement of formation itself. For the
symmetric amplitude damping, when there is not phase damping and $N\geq 5,$ $%
E_f^{+}(\rho )$ is very close to $E_r(\rho ).$

The distillable entanglement is lower bounded by the coherent information
(hashing inequality). The coherent information of the state is
\begin{eqnarray}
I_c(\rho ) &=&-c_{N0}\log _2c_{N0}-(\sum_{m=1}^Nc_{0m})\log
_2(\sum_{m=1}^Nc_{0m}) \\
&&+\sum_{m=1}^{N-1}c_{0m}\log _2(c_{0m})+c_{+}\log _2c_{+}+c_{-}\log _2c_{-}.
\nonumber
\end{eqnarray}
When only phase damping is considered, that is $\Gamma _1=\Gamma _2=0,$ we
have $c_{0m}=0$ for all $m<N,$ $c_{N0}=\frac 12,c=\frac 12e^{-N^2\overline{%
\gamma }t}.$ The coherent information will be $I_c(\rho )=1-H_2(\frac
12+\frac 12e^{-N^2\overline{\gamma }t}).$ Meanwhile the relative entropy of
entanglement will also be $E_r(\rho )=1-H_2(\frac 12+\frac 12e^{-N^2%
\overline{\gamma }t}).$ Because distillable entanglement $E_d(\rho )$ is
upper bounded by the relative entropy of entanglement, we have $I_c(\rho
)\leq E_d(\rho )\leq E_r(\rho )$. Now $I_c(\rho )=E_r(\rho )$, thus for the
situation only phase damping, the distillable entanglement is:
\begin{equation}
E_d(\rho )=1-H_2(\frac 12+\frac 12e^{-N^2\overline{\gamma }t}).
\end{equation}

\section{Performance of damped state in entanglement enhanced phase
measurement}

In the entanglement enhanced phase measurement, the measurement operator is $%
A=\left| N0\right\rangle \left\langle 0N\right| +\left| 0N\right\rangle
\left\langle N0\right| $\cite{Mitchell}$.$ We suppose the state undergo the
phase shift just before the measurement apparatus. The state is modified by
the phase shift to $\rho _\varphi =c_{00}\left| 00\right\rangle \left\langle
00\right| +\sum_{m=1}^N(c_{m0}\left| m0\right\rangle \left\langle m0\right|
+ $ $c_{0m}\left| 0m\right\rangle \left\langle 0m\right| )$ $+c(e^{iN\varphi
}\left| N0\right\rangle \left\langle 0N\right| +e^{-iN\varphi }\left|
0N\right\rangle \left\langle N0\right| ).$ It should be noted that the
relative entropy of entanglement is not changed by the phase shift\cite
{Chen1}. Thus $\left\langle A\right\rangle =Tr\rho _\varphi A=$ $2c\cos
N\varphi =$ $e^{-N\overline{\Gamma }t-N^2\overline{\gamma }t}\cos N\varphi ,$
$\Delta A=\sqrt{\left\langle A^2\right\rangle -\left\langle A\right\rangle ^2%
},$ the phase deviation is
\begin{equation}
\Delta \varphi =\frac{\Delta A}{\left| \partial \left\langle A\right\rangle
/\partial \varphi \right| }=\frac{\sqrt{\frac 12(e^{-N\Gamma
_1t}+e^{-N\Gamma _2t})-e^{-2N\overline{\Gamma }t-2N^2\overline{\gamma }%
t}\cos ^2N\varphi }}{Ne^{-N\overline{\Gamma }t-N^2\overline{\gamma }t}\left|
\sin N\varphi \right| }.
\end{equation}
The minimal $\Delta \varphi $ will be at $\varphi =(2k+1)\frac \pi {2N}.$
Thus the bset phase measurement precision will be
\begin{equation}
(\Delta \varphi )_{best}=\frac 1Ne^{N^2\overline{\gamma }t}\sqrt{\frac
12(e^{N\Gamma _1t}+e^{N\Gamma _2t})}.
\end{equation}

We may adopt another strategy to enhance the phase measurement precision
with the mixed entangled state at hand. In this strategy, the damped state $%
\rho $ is distilled to \textit{noon} state, then we use the new \textit{noon}
state for measurement. The successful probability of distilling a \textit{%
noon} state from the damped state is characterized by the distillable
entanglement of $\rho .$ Thus we use \textit{noon} state in the measurement
at a probability of $p=E_d(\rho )$, the phase deviation is the Heisenberg
limit $(\Delta \varphi )_q=\frac 1N$; In a probability of $1-E_d(\rho )$ we
have no quantum entanglement to enhance the measurement, the phase deviation
will be the classical shot noise limit $(\Delta \varphi )_c=\frac 1{\sqrt{N}%
}.$ The mixture of the two kind of measurements will have the phase
deviation: $(\Delta \varphi )_d=\sqrt{p(\Delta \varphi )_q^2+(1-p)(\Delta
\varphi )_c^2},$ (this may derived from the fact that the mixture
probability distribution function $f_d(\varphi )=pf_q(\varphi
)+(1-p)f_c(\varphi ),$ with $f_q(\varphi )$ and $f_c(\varphi )$ are
probability distribution function of entanglement enhanced phase measurement
and classical measurement, the two kinds of measurement have the same mean).
In the distillation strategy, the phase deviation will be
\begin{equation}
(\Delta \varphi )_d=\sqrt{E_d/N^2+(1-E_d)/N}.
\end{equation}

We can compare the performances of the direct application of damped state to
phase measurement and the distillation then measurement strategy. Let us
firstly consider the situation of phase damping alone. We have
\begin{equation}
\frac{(\Delta \varphi )_d^2}{(\Delta \varphi )_{best}^2}=[1+(N-1)H_2(\frac
12+\frac 12e^{-N^2\overline{\gamma }t})]e^{-2N^2\overline{\gamma }t}.
\end{equation}
\begin{figure}[tbp]
\includegraphics[height=3in]{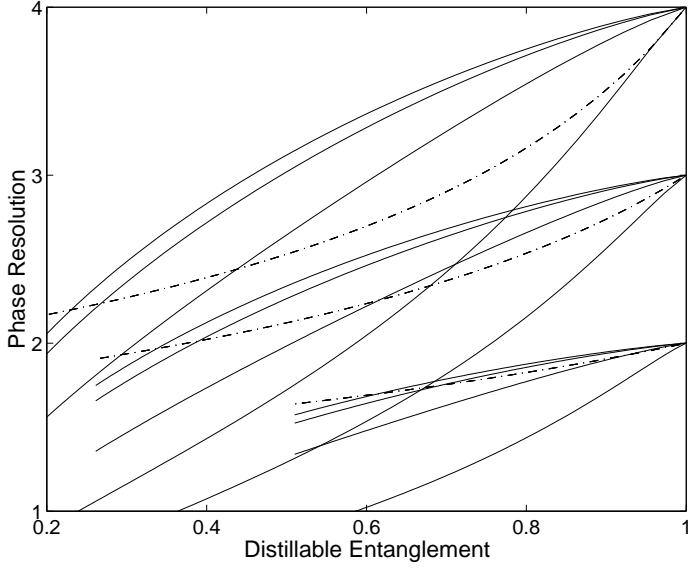}
\caption{The performance of state with phase damping alone in entanglement
enhanced measurement. The phase resolution is defined as the inverse of
phase deviation. The line groups from top to bottom are for $N=4,3,2$
respectively. In each group the solid lines from top to bottom are $1/\Delta
\varphi $ for $\varphi =(1,3/4,1/2,1/4)\pi /(2N)$ respectively, the dashed
line is for $1/(\Delta \varphi )_d$. }
\end{figure}

As indicated in figure 1, the performance of direct application is better
when the phase $\varphi $ under measurement is near $(2k+1)\pi /(2N)$. While
distillation strategy is better when the phase $\varphi $ is far from these
values.

In the amplitude damping situation, distillable entanglement is upper and
lower bounded by the relative entropy of entanglement and the coherent
information respectively. It is followed that the phase deviation is also
upper and lower bounded. We have $(\Delta \varphi )_{dl}\leq (\Delta \varphi
)_d\leq (\Delta \varphi )_{du}$ with $(\Delta \varphi )_{dl}=\sqrt{%
E_r/N^2+(1-E_r)/N}$ and $(\Delta \varphi )_{du}=\sqrt{I_c/N^2+(1-I_c)/N}.$
In figure 2, we calculate the upper and lower bounds of resolution for the
distillation strategy in situation of symmetric amplitude damping alone. We
can see that the best resolution of the direct application of damped state
is better than the distillation strategy.

\begin{figure}[tbp]
\includegraphics[height=3in]{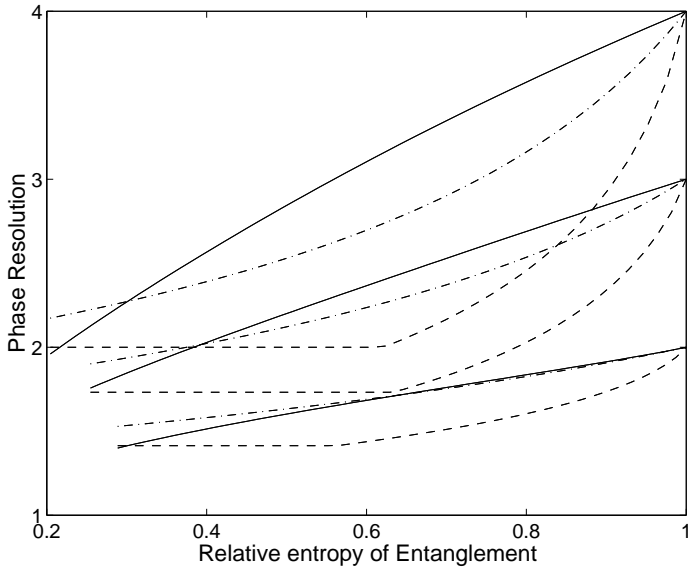}
\caption{The performance of state with amplitude damping alone in
entanglement enhanced measurement. The phase resolution is defined as the
inverse of phase deviation. The line groups from top to bottom are for $%
N=4,3,2$ respectively. In each group the solid line is for $1/(\Delta
\varphi )_{best}$, the dotdash line is for $1/(\Delta \varphi )_{dl}$,the
dashed line is for $1/(\Delta \varphi )_{du}$. }
\end{figure}

\section{Conclusions}

The master equation of quantum continuous variable system is sovled in the
case of simultaneous amplitude damping of vacuum environment and phase
damping. When the initial state is a \textit{noon }state, the exact
expression of time dependent solution of density operator is obtained via
the characteristic function method. An analytical formula is given for the
relative entropy of entanglement of the damped state when the two modes of
the \textit{noon }state undergo the same amount of amplitude damping. In the
asymmetric amplitude damping, the the relative entropy of entanglement can
be calculated by numerically solving a group of algebraic equations. In the
situation of phase damping alone, the exact distillable entanglement is
given, which enables the comparison of two strategies of applying the damped
state in phase measurement possible. When amplitude damping is present, we
calculate the relative entropy of entanglement and coherent information of
the damped state. We use these to specify the upper and lower bounds of the
distillable entanglement.

The performance of direct application of the damped state in phase
measurement is better than that of firstly distilling the damped state then
applying it in measurement when the phase under estimation is near $%
(2k+1)\pi /(2N)$. While distillation strategy is better when the phase $%
\varphi $ is far from these values.

\section*{Acknowledgment}

Funding by Zhejiang Province Natural Science Foundation (Grant No.
RC104265), AQSIQ of China (Grant No. 2004QK38) and the National Natural
Science Foundation of China (Grant No. 10575092, No. 10347119) are
gratefully acknowledged.

\section*{Appendix: Proof of the extremal state}

We prove that $\sigma ^{*}$ is extremal by the fact that local minimum is
also the global minimum when it is in regard to the relative entropy of
entanglement\cite{Vedral2}. Hence we only need to prove that $\sigma ^{*}$
is the local minimal state. Let $f(x,\sigma ^{*},\sigma )=S(\rho \left\|
(1-x)\sigma ^{*}+x\sigma \right. )$ be the relative entropy of a state
obtained by moving from $\sigma ^{*}$ towards some $\sigma $. The derivative
of $f$ will be \cite{Rehacek}\cite{Vedral2}
\[
\frac{\partial f}{\partial x}\left( 0,\sigma ^{*},\sigma \right)
=\int_0^\infty ((\sigma ^{*}+t)^{-1}\rho (\sigma ^{*}+t)^{-1}\delta \sigma
)dt=TrB\delta \sigma ,
\]
where we denote $(1-x)\sigma ^{*}+x\sigma =\sigma ^{*}-\delta \sigma ,$ and
the operator $B$ has the following matrix elements in the eigenbasis $%
\{\left| \chi _n\right\rangle \}$ of $\sigma ^{*}$:
\[
B_{mn}^\chi =\left\langle \chi _m\right| B\left| \chi _n\right\rangle =\frac{%
\log \chi _n-\log \chi _m}{\chi _n-\chi _m}\left\langle \chi _m\right| \rho
\left| \chi _n\right\rangle .
\]
And when $\chi _m=\chi _n$, the corresponding coefficient should be replaced
with the limit value of $\chi _n^{-1}.$

We should prove that for any separable state $\chi $,
\[
\frac{\partial f}{\partial x}\left( 0,\sigma ^{*},\sigma \right) =TrA\delta
\sigma =TrA(\sigma ^{*}-\sigma )\geq 0.
\]
where $TrB\sigma ^{*}=1$\cite{Rehacek}, But any $\sigma \in \mathcal{D}$
(separable state set) can be written in the form of $\sigma =\sum_ip_i\left|
\alpha ^i\beta ^i\right\rangle \left\langle \alpha ^i\beta ^i\right| $ and
so $\frac{\partial f}{\partial x}\left( 0,\sigma ^{*},\sigma \right)
=\sum_ip_i\frac{\partial f}{\partial x}\left( 0,\sigma ^{*},\left| \alpha
^i\beta ^i\right\rangle \left\langle \alpha ^i\beta ^i\right| \right) $, The
problem is reduced to prove that for any normalized pure state $\left|
\alpha \beta \right\rangle \left\langle \alpha \beta \right| ,$%
\begin{equation}
\left\langle \alpha \beta \right| B\left| \alpha \beta \right\rangle \leq 1.
\tag{A1}  \label{wave5}
\end{equation}
We here prove the situation of symmetric amplitude damping system, the more
general proof for corresponding two qubits system was already been found\cite
{Chen1}, and the proof of asymmetric amplitude damping system is similar to
that of two qubits system. For symmetric amplitude damping , the operator
\begin{eqnarray*}
B &=&\frac{c_{00}}{d_{00}}\left| 00\right\rangle \left\langle 00\right|
+\sum_{m=1}^{N-1}(\left| m0\right\rangle \left\langle m0\right| +\left|
0m\right\rangle \left\langle 0m\right| ) \\
&&+\frac 12(\frac{c_{+}}{d_{+}}+\frac{c_{-}}{d_{-}})(\left| N0\right\rangle
\left\langle N0\right| +\left| 0N\right\rangle \left\langle 0N\right|
)+\frac 12(\frac{c_{+}}{d_{+}}-\frac{c_{-}}{d_{-}})(\left| N0\right\rangle
\left\langle 0N\right| +\left| 0N\right\rangle \left\langle N0\right| ).
\end{eqnarray*}
Denote $\left| \alpha \right\rangle =\sum_m\alpha _m\left| m\right\rangle ,$
thus $\left\langle \alpha \beta \right| B\left| \alpha \beta \right\rangle =%
\frac{c_{00}}{d_{00}}\left| \alpha _0\beta _0\right|
^2+\sum_{m=1}^{N-1}(\left| \alpha _m\beta _0\right| ^2+\left| \alpha _0\beta
_m\right| ^2)$ $+\frac 12(\frac{c_{+}}{d_{+}}+\frac{c_{-}}{d_{-}})(\left|
\alpha _0\beta _N\right| ^2+\left| \alpha _N\beta _0\right| ^2)+\frac 12(%
\frac{c_{+}}{d_{+}}-\frac{c_{-}}{d_{-}})(\alpha _N\beta _0\alpha _0^{*}\beta
_N^{*}+\alpha _0\beta _N\alpha _N^{*}\beta _0^{*}).$ Let $%
K_1=1-\sum_{m=1}^{N-1}\left| \alpha _m\right| ^2$ ,$K_2=1-\sum_{m=1}^{N-1}%
\left| \beta _m\right| ^2,$ (with $0\leq K_1,K_2\leq 0),$ and $\alpha _0=%
\sqrt{K_1}\cos \theta _1,\alpha _N=\sqrt{K_1}\sin \theta _1e^{i\eta _1},$ $%
\beta _0=\sqrt{K_2}\cos \theta _2e^{i\eta _2},$ $\beta _N=\sqrt{K_2}\sin
\theta _2e^{i\eta _3}.$ Denote $\eta =\eta _2-\eta _1-\eta _3$, after
maximization on $\eta $, we have
\begin{eqnarray*}
\left\langle \alpha \beta \right| B\left| \alpha \beta \right\rangle &=&%
\frac{c_{00}}{d_{00}}K_1\cos ^2\theta _1\cos ^2\theta _2+(1-K_1)K_2\cos
^2\theta _2+(1-K_2)K_1\cos ^2\theta _1 \\
&&+\frac{c_{+}}{2d_{+}}K_1K_2\sin ^2(\theta _1+\theta _2)+\frac{c_{-}}{2d_{-}%
}K_1K_2\sin ^2(\theta _1-\theta _2).
\end{eqnarray*}
Suppose the extremal value is achieved at some $K_1\neq 0,1,$ by derivative
on $K_1,$ we obtain the supposed (may not exist) extremal value $%
\left\langle \alpha \beta \right| B\left| \alpha \beta \right\rangle
=K_2\cos ^2\theta _2\leq 1.$ What left is to verify that when $K_1,K_2=0$ or
$1,$ $\left\langle \alpha \beta \right| B\left| \alpha \beta \right\rangle $
will not exceeds $1$. The nontrivial situation is $K_1=K_2=1,$ thus
\[
\left\langle \alpha \beta \right| B\left| \alpha \beta \right\rangle =\frac{%
c_{00}}{4d_{00}}(\cos \phi _1+\cos \phi _2)^2+\frac{c_{+}}{2d_{+}}\sin
^2\phi _1+\frac{c_{-}}{2d_{-}}\sin ^2\phi _2,
\]
where $\phi _{1,2}=\theta _1\pm \theta _2.$ By using the fact that $\frac{%
c_{00}}{d_{00}}(\frac{c_{+}}{d_{+}}+\frac{c_{-}}{d_{-}})=2\frac{c_{+}c_{-}}{%
d_{+}d_{-}},$ We can obtain the maximum value as
\[
\left\langle \alpha \beta \right| B\left| \alpha \beta \right\rangle
_m=\frac 12(\frac{c_{+}}{d_{+}}+\frac{c_{-}}{d_{-}})=1.
\]
Hence inequality (\ref{wave5}) is proved. So that $\sigma ^{*}$ is the
extremal state that minimizes the relative entropy.


\begin{thebibliography}{99}
\bibitem{Holland}  M. J. Holland and K. Burnett, Phys. Rev. Lett. \textbf{71}%
, 1355(1993).

\bibitem{Bollinger}  J. J. Bollinger, W. M. Itano, D. J. Wineland, and D.
J.Heinzen, Phys. Rev. A \textbf{54}, R4649 (1996).

\bibitem{Dowling}  J. P. Dowling, Phys. Rev. A \textbf{57}, 4736 (1998).

\bibitem{DAngelo}  M. D'Angelo, M.V. Chekhova, and Y. Shih, Phys. Rev. Lett.
\textbf{87}, 013602 (2001).

\bibitem{Mitchell}  M. W. Mitchell, J. S. Lundeen, and A. M. Steinberg,
Nature \textbf{429}, 161 (2004).

\bibitem{Walther}  P. Walther, J.-W. Pan, M. Aspelmeyer, R. Ursinand, S.
Gasparoni, and A. Zeilinger, Nature \textbf{429}, 158 (2004).

\bibitem{Campos}  R. A. Campos, C. C. Gerry, and A. Benmoussa, Phys. Rev. A
68, 023810 (2003).

\bibitem{Boto}  A.N. Boto, P. Kok, D.S. Abrams, S.L. Braunstein, C.P.
Williams, and J.P. Dowling, Phys. Rev. Lett. \textbf{85}, 2733 (2000); G.S.
Agarwal, R.W. Boyd, E.M. Nagasako, and S.J. Bentley, Phys. Rev. Lett.
\textbf{86}, 1389 (2001).

\bibitem{Huelga}  S. F. Huelga, C. Macchiavello, T. Pellizzari, A. K. Ekert,
M. B. Plenio, and J. I. Cirac, Phys. Rev. Lett. \textbf{79}, 3865 (1997).

\bibitem{Kinsler}  P. Kinsler and P. D. Drummond, Phys. Rev. A 43, 6194
(1991).

\bibitem{Lindblad}  G. Lindblad, Commun. Math. Phys. 48, 119 (1976).

\bibitem{Walls}  D. Walls and G. Milburn, \textit{Quantum optics }(Springer
Verlag, Berlin, 1994).

\bibitem{Chen}  X.Y. Chen, \textit{Phys. Rev. A} \textbf{73}, 022307 (2006).

\bibitem{Perelomov}  A. Perelomov, \textit{Generalized Coherent states},
Springer Verlag, Berlin (1986).

\bibitem{Chen0}  X.Y. Chen, \textit{Phys. Rev. A} \textbf{71}, 062320 (2005).

\bibitem{Chen1}  X.Y. Chen, L. M. Meng, L. Z. Jiang, and X. J. Li, \textit{%
Chin.Phys. Lett} \textbf{22}, 2755 (2005).

\bibitem{Vedral}  V.Vedral, M. B. Plenio \textit{Phys. Rev. A} \textbf{57,}%
1619 (1998).

\bibitem{Vedral2}  V. Vedral, M. B. Plenio , K. Jacobs and P. L. Knight,
\textit{Phys. Rev. A} \textbf{56}, 4452 (1997).

\bibitem{Rehacek}  J. \v {R}eh\'{a}\v {c}ek and Z. Hradil, Phys. Rev. Lett.
\textbf{90}, 127904 (2003).
\end{thebibliography}
\end{document}